\documentclass{ws-mpla}

\begin{document}

\markboth{Ulhoa, Santos and Amorim}
{On the Energy-Momentum Flux in G\"{o}del-type Models}

\catchline{}{}{}{}{}

\title{On the Energy-Momentum Flux in G\"{o}del-type Models}

\author{\footnotesize S. C. Ulhoa}

\address{Faculdade Gama, Universidade de Bras\'{i}lia, 72444-240, Setor Leste (Gama), Bras\'{i}lia, DF, Brazil. \\
sc.ulhoa@gmail.com}

\author{A. F. Santos}

\address{Instituto de F\'isica, Universidade Federal de Mato Grosso,
78060-900, Cuiab\'a, MT, Brasil.\\alesandroferreira@fisica.ufmt.br }

\author{R.G.G. Amorim}

\address{Faculdade Gama, Universidade de Bras\'{i}lia, 72444-240, Setor Leste (Gama), Bras\'{i}lia, DF, Brazil.\\ronniamorim@gmail.com }

\maketitle

\pub{Received (Day Month Year)}{Revised (Day Month Year)}

\begin{abstract}
In this paper we work in the context of Teleparallelism Equivalent to General Relativity (TEGR) in order to construct the energy-momentum flux for G\"odel-type solutions of Einstein's equations. We use an stationary observer, which is settled by the tetrad choice, to obtain the gravitational pressure for each direction of space in cartesian coordinates. Then we write down the total pressure for each direction in terms of the pressure of the fluid, thus we are able to identify role of the gravitational pressure.

\keywords{Teleparallelism; Torsion Tensor; G\"odel Universe.}
\end{abstract}

\ccode{PACS Nos.: 04.20-q; 04.20.Cv; 02.20.Sv}

\section{Introduction}
\noindent

The idea of a rotating universe was first proposed by G. Gamow in 1946~\cite{Gamow}. Not long after that, in 1949 K. G\"{o}del obtained the first solution of Einstein's  equations for a rotating universe \cite{Godel}. This solution is stationary, spatially homogeneous and it displays cylindrical symmetry. In addition G\"{o}del's solution has a highly nontrivial feature:  it can violate causality which implies the possibility of closed timelike curves (CTC). In recent years there has been great interest on G\"odel-type metrics, for instance the problem of causality was examined with more details in the references \refcite{Reb,Reb1,Reb2}. A more close analysis of G\"odel-type metrics allows us to establish three different classes of solutions. These solutions are characterized by the following possibilities:  (i)  there is no CTCs, (ii)  there is an infinite sequence of alternating causal and noncausal regions, and (iii)  there is only one noncausal region. The consistency of these solutions was checked for other gravitational models such as Chern-Simons modified gravity \cite{CS,CS1}, $f(R)$ gravity \cite{Santos} and Horava-Lifshitz gravity \cite{Horava}. The compatibility of the G\"{o}del metric with the usual Einstein-Aether theory have been discussed in the reference \refcite{Gurses} as well.

An interesting generalization of G\"{o}del's work which is also a G\"{o}del-type metric, is the model proposed by Obukhov \cite{Obukhov}, there the author propose a universe rotating and expanding. The Obukhov's model was also studied in the context of teleparallel gravity, see \refcite{Vargas,Adellane} and the references therein. In this article we shall study the total energy-momentum flux for a model of universe which presents just rotation, since the energy-momentum for a perfect fluid will be used, in the realm of Teleparallelism Equivalent to General Relativity.

Teleparallel gravity and General Relativity (GR) are identical under the point of view of dynamics, which means the equivalence of field equations. It does not mean whatsoever the same predictions. The relation between TEGR and GR is roughly speaking quite the same between Newtonian formalism and Lagrangian or Hamiltonian formalism. Despite the fact that the equations of motion are the same, in Lagrangian formalism it is possible to understand conserved quantities (Noether theorem) in a way not possible in Newtonian formalism. It worths to recall that both theories share the same metric tensor. As a matter of fact the metric tensor can be determined by means Einstein equations, then it can be used to construct the quantities in the realm of Teleparallelism Equivalent to General Relativity (TEGR), such as the gravitational energy, once the reference frame is chosen. Because of this we can use for example the Schwarzschild solution (Kerr solution or any available solution) to construct the quantities predicted by TEGR. In the metrical formulation it is not known how to obtain an expression like the gravitational energy-momentum $t^{\mu\nu}$, in terms of the tetrad field, which is a tensor under coordinates transformation. Actually it is impossible to derive a tensorial expression that represents the gravitational energy from Einstein equations in terms of the metric tensor. Therefore GR is neither a final nor an absolute theory, in this sense it needs at any point an improvement. We believe that TEGR can provide such an enhancement.

This article is structured as follows. In section \ref{tel}  we briefly introduce teleparallel gravity. We present expressions for the total energy-momentum flux which takes into account both the gravitational and matter field fluxes. We also recall the definition of the total pressure which will be used latter. In section \ref{godel} we write down the line element of the G\"odel-type Universe and calculate the total energy-momentum flux. We obtain that the total pressure is equal to the pressure of the perfect fluid plus something else which can be viewed as the gravitational pressure. Finally in the last section we present some concluding remarks.

\bigskip
Notation: space-time indices $\mu, \nu, ...$ and SO(3,1) indices $a,
b, ...$ run from 0 to 3. Time and space indices are indicated
according to $\mu=0,i,\;\;a=(0),(i)$. The tetrad field is denoted by
$e^a\,_\mu$ and the determinant of the tetrad field is represented
by $e=\det(e^a\,_\mu)$. In addition we adopt units where
$G=c=1$, unless otherwise stated.\par
\bigskip

\section{Teleparallelism Equivalent to General Relativity (TEGR)\label{tel}}
\noindent

In this section we will recall the main ideas concerned to teleparallel gravity which is centered, as its dynamical variables, in the tetrad field instead of the metric tensor of general relativity. The relation between the tetrad field and the metric tensor is established by $g_{\mu\nu}=e_{a\mu}e^a\,_{\nu}$. Thus the tetrad field ensures naturally the Lorentz transformations as distinct features from coordinate transformations.  We will start with a Weitzenb\"ock geometry and show its connection to the Riemannian one.

Let us first consider a manifold to which is assigned the connection
$\Gamma_{\mu\lambda\nu}=e^{a}\,_{\mu}\partial_{\lambda}e_{a\nu}$,
which is called  Cartan connection~\cite{Cartan}. Such a manifold defines the Weitzenb\"ock geometry in the same sense that with the help of the Christoffel symbols is possible to define a Riemannian manifold. Differently from the Christoffel symbols whose torsion vanishes identically, the Cartan connection yields a vanishing curvature. The torsion tensor constructed from the Cartan connection, as its skew-symmetric part, is given by

\begin{equation}
T^{a}\,_{\lambda\nu}=\partial_{\lambda} e^{a}\,_{\nu}-\partial_{\nu}
e^{a}\,_{\lambda}\,. \label{4}
\end{equation}

The Christoffel symbols, ${}^0\Gamma_{\mu \lambda\nu}$, can be expressed in terms of the tetrads which gives rise to the following identity

\begin{equation}
\Gamma_{\mu \lambda\nu}= {}^0\Gamma_{\mu \lambda\nu}+ K_{\mu
\lambda\nu}\,, \label{2}
\end{equation}
where $K_{\mu \lambda\nu}$ is the contortion tensor. It can be given in terms of the
torsion tensor constructed from the  Cartan connection which reads

\begin{eqnarray}
K_{\mu\lambda\nu}&=&\frac{1}{2}(T_{\lambda\mu\nu}+T_{\nu\lambda\mu}+T_{\mu\lambda\nu})\,.\label{3}
\end{eqnarray}
The torsion tensor is $T_{\mu \lambda\nu}=e_{a\mu}T^{a}\,_{\lambda\nu}$.

The scalar curvature of a Riemannian manifold obtained from  ${}^0\Gamma_{\mu \lambda\nu}$, in terms of the tetrad field, is designated by $R(e)$. Thus using (\ref{2}) and (\ref{3}), it leads to

\begin{equation}
eR(e)\equiv -e({1\over 4}T^{abc}T_{abc}+{1\over
2}T^{abc}T_{bac}-T^aT_a)+2\partial_\mu(eT^\mu)\,.\label{5}
\end{equation}
It should be noted that the left hand-side of (\ref{5}) is the Hilbert-Einstein lagrangian density since the determinant of the tetrad is equal to the square root of the negative determinant of the metric tensor, $e=\sqrt{-g}$. It worths to note that a total divergence in the lagrangian does not bestow anything new to the field equations. Thus the equivalence between TEGR and GR can be settled by the use of the following  Teleparallel Lagrangian density

\begin{eqnarray}
\mathfrak{L}(e_{a\mu})&=& -\kappa\,e\,({1\over 4}T^{abc}T_{abc}+
{1\over 2} T^{abc}T_{bac} -T^aT_a) -\mathfrak{L}_M\nonumber \\
&\equiv&-\kappa\,e \Sigma^{abc}T_{abc} -\mathfrak{L}_M\;, \label{6}
\end{eqnarray}
where $\kappa=1/(16 \pi)$, $\mathfrak{L}_M$ is the Lagrangian
density of matter fields and $\Sigma^{abc}$ is given by

\begin{equation}
\Sigma^{abc}={1\over 4} (T^{abc}+T^{bac}-T^{cab}) +{1\over 2}(
\eta^{ac}T^b-\eta^{ab}T^c)\;, \label{7}
\end{equation}
with $T^a=e^a\,_\mu T^\mu$.

The field equations can be derived from (\ref{6}) by means a  variational derivative of the Lagrangian density with
respect to $e^{a \mu}$ which reads

\begin{equation}
e_{a\lambda}e_{b\mu}\partial_\nu(e\Sigma^{b\lambda \nu})-
e(\Sigma^{b \nu}\,_aT_{b\nu \mu}- {1\over
4}e_{a\mu}T_{bcd}\Sigma^{bcd}) \;= {1\over {4\kappa}}eT_{a\mu}\,,
\label{8}
\end{equation}
where $T_{a\mu}=e_{a}\,^{\lambda }T_{\mu
\lambda}=\frac{1}{e}\frac{\delta {\mathcal{L}}_{M}}{\delta e^{a\mu
}}$ is the energy-momentum tensor of matter fields. It is possible
to show by explicit calculations the equivalence of eq. (\ref{8})
and Einstein equations~\cite{maluf:335} which is not a surprise at all given the equivalence between the lagrangian density of TEGR and GR.

It is possible to reexpress the field equations as

\begin{equation}
\partial_\nu(e\Sigma^{a\lambda\nu})={1\over {4\kappa}}
e\, e^a\,_\mu( t^{\lambda \mu} + T^{\lambda \mu})\;, \label{10}
\end{equation}
where $t^{\lambda\mu}$ is defined by

\begin{equation}
t^{\lambda \mu}=\kappa(4\Sigma^{bc\lambda}T_{bc}\,^\mu- g^{\lambda
\mu}\Sigma^{bcd}T_{bcd})\,. \label{11}
\end{equation}
The symmetry of  $\Sigma^{a\lambda\nu}$ leads to

\begin{equation}
\partial_\lambda(et^{a\lambda}+eT^{a\lambda})=0\,\label{13}
\end{equation}
which is a local conservation law for the total (gravitational and matter fields) energy-momentum tensor. The above equation can be expanded which reads

$$
{d\over {dt}} \int_V d^3x\,e\,e^a\,_\mu (t^{0\mu} +T^{0\mu})
=-\oint_S dS_j\, \left[e\,e^a\,_\mu (t^{j\mu} +T^{j\mu})\right] \,.
$$
Thus $t^{\lambda \mu}$ is interpreted as the
energy-momentum tensor of the gravitational field~\cite{PhysRevLett.84.4533,maluf2}.
Therefore, it is possible to define the total energy-momentum contained in a three-dimensional
volume $V$ of space as

\begin{equation}
P^a = \int_V d^3x \,e\,e^a\,_\mu(t^{0\mu}+ T^{0\mu})\,. \label{14}
\end{equation}
Such an expression is dependent on the choice of the reference frame, since it is a vector under Lorentz transformations. The energy-momentum vector is also invariant under coordinates transformations which means the same physical predictions for each coordinate system used. These features are present in the Special Theory of Relativity and there is no good reason to give up them when dealing with a theory of gravitation.

Let us define the gravitational energy-momentum flux as

\begin{equation}
\Phi^a_g=\oint_S dS_j\,
\, (e\,e^a\,_\mu t^{j\mu})\,,
\label{15}
\end{equation}
and

\begin{equation}
\Phi^a_m=\oint_S dS_j\,
\,( e\,e^a\,_\mu T^{j\mu})\,,
\label{16}
\end{equation}
as the energy-momentum flux of matter fields. Thus

\begin{eqnarray}
{{dP^a}\over {dt}}&=&-\left(\Phi^a_g+\Phi^a_m\right)\nonumber\\
&=&-4k\oint_S dS_j\,
\partial_\nu(e\Sigma^{a j\nu})\,.
\label{17}
\end{eqnarray}
Then if we restrict our attention to the spatial indices in the last equation, we have
\begin{equation}
{{dP^{(i)}}\over {dt}}= -\oint_S dS_j\, \phi^{(i)j} \,,
\label{18}
\end{equation}
where

\begin{equation}
\phi^{(i)j}=4k\partial_\nu(e\Sigma^{(i)j\nu}) \,.
\label{19}
\end{equation}
Equation (\ref{18}) represents the derivative of the momentum with respect to the time which is the force acting on the system. It should be noted also that $dS_j$ is an element of area. Therefore  $-\phi^{(i)j}$ represents the pressure along the $(i)$ direction, over and element of area oriented along
the $j$ direction.

\section{The G\"odel-Type Space-Time\label{godel}}
\noindent

In this section we will study the role of pressure in a universe described by the G\"odel-type space-time which is defined by the following line element

\begin{equation}
ds^2=-dt^2+2\alpha(t)\sqrt{\sigma}{\rm e}^{mx}dtdy+\alpha(t)^2\left(dx^2+k'{\rm e}^{2mx}dy^2+dz^2\right)
\,.\label{17}
\end{equation}

In such a system the matter fields will be described by a perfect fluid with energy-momentum given by
\begin{equation}
T^{\mu\nu}=(\epsilon+p)U^{\mu}U^{\nu}+pg^{\mu\nu}\,,\nonumber
\end{equation}
where $\epsilon$, $p$ and $U^\mu$ are the fluid energy density, the fluid
pressure and the fluid four-velocity field,
respectively~\cite{Dinverno}.

The tetrad is chosen like
\begin{equation}
e^a\,_\mu= \left[ \begin {array}{cccc} 1&0&-\alpha \left( t \right)\, \sqrt {\sigma}\,{{\rm e}^
{mx}}&0 \\\noalign{\medskip}0&\alpha \left( t \right) &0&0
\\\noalign{\medskip}0&0&\sqrt {k'+\sigma}\,\,\alpha \left( t \right)\, {{\rm e}
^{mx}}&0\\\noalign{\medskip}0&0&0&\alpha \left( t \right)
\end {array} \right]\,,\label{tetrada}
\end{equation}
in order to be adapted to a stationary reference frame. There are essentially two ways to establish a tetrad field, one of them is based on the acceleration tensor introduced by Mashhoon~\cite{mashhoon}, on the other hand it is possible to associate the components of the inverse tetrad, $e_{(0)}\,^\mu$, to the field velocity of the observer. Thus the above tetrad is adapted to an observer with field velocity $(1,0,0,0)$ and hence stationary.

The non-vanishing components of the torsion calculated from the tetrad field (\ref{tetrada}) are

\begin{eqnarray}
T_{002}=\dot{\alpha}(t)\sqrt {\sigma}{{\rm e}^{mx}}\,, &\qquad& T_{012}=\alpha \left( t \right) \sqrt {\sigma}m{{\rm e}^{mx}}\,,\nonumber\\
T_{101}= \alpha(t)\dot{\alpha}(t)\,,&\qquad& T_{202}= k'{{\rm e}^{2\,mx}}\alpha(t)\dot{\alpha}(t)\,,\nonumber\\
T_{212}=\left( \alpha \left( t \right)  \right) ^{2}k'\,m{{\rm e}^{2\,mx}}\,,&\qquad&T_{303}=\alpha(t)\dot{\alpha}(t)\,,
\end{eqnarray}
where the dot means temporal derivative.

The dynamics of $\alpha(t)$ is given by the field equations which reads

\begin{eqnarray}
\left[\frac{3(\dot{\alpha}(t))^2-m^2(k^{\prime}+\sigma/4)}{\alpha(t)^2(k^{\prime}+\sigma)}\right]&=&8\pi\epsilon\,,\nonumber\\
\left[\frac{k^{\prime}(\left( \dot{\alpha}(t)
 \right) ^{2}+2\alpha \left( t \right) \ddot{\alpha}\left( t \right))-m^2(k^{\prime}+3\sigma/4)}{\alpha(t)^2(k^{\prime}+\sigma)}\right]&=&-8\pi p\,.\nonumber\\
\end{eqnarray}
These equations reduce to FRW equations if we choose $m=\sigma=0$ and $k'=1$.

The relevant components of $\Sigma^{a\mu\nu}$ to evaluate the energy flux are

\begin{eqnarray}
e\,\Sigma^{(0)01}&=&-\frac{1}{4}\,\left[{\frac {\alpha \left( t \right) {{\rm e}^{mx}}m \left( \sigma+2
\,k' \right) }{\sqrt {k'+\sigma}}}\right]\,,\\
e\,\Sigma^{(0)02}&=&-\left(\frac { \sqrt {\sigma} }{\sqrt {k'+\sigma}}\right)\,\alpha(t)\dot{\alpha}(t)\,,
\end{eqnarray}
then after some algebraic manipulations we get

\begin{equation}
\frac{dE}{dt}=-\left(\frac{L^2}{16\pi}\right)\left(\frac{2k^{\prime}+\sigma}{\sqrt{k^{\prime}+\sigma}}\right)me^{mL}\,\dot{\alpha}(t)\,.\label{flux}
\end{equation}
In order to find such an energy flux, we have supposed a squared surface of integration with area equal to $L^2$. Indeed equation (\ref{flux}) represents the energy flux through an open surface of area $L^2$.

Now let us calculate the momentum flux through an open surface of unitary area, for this purpose we need the following components of $\phi^{(i)j}$

\begin{eqnarray}
\phi^{(1)1}&=&-4k\left[\frac {  \left( \dot{\alpha}(t) \right) ^{2}+\alpha \left( t \right) \ddot{\alpha}\left( t \right)}{\sqrt {k'+\sigma}}\right]\, k'{\rm e}^{mx}\,\\
\phi^{(1)2}&=&k\,\left[\frac {\dot{\alpha}(t)
\sqrt{\sigma}\,m}{\sqrt {k'+\sigma}}\right]\,,\\
\phi^{(2)1}&=&-k\, \dot{\alpha}(t) \sqrt {
\sigma}\,m{{\rm e}^{mx}}\,,\\
\phi^{(2)2}&=&-4k\left[\left( \dot{\alpha}(t)
 \right) ^{2}+\alpha \left( t \right) \ddot{\alpha}\left( t \right)\right]\,,\\
\phi^{(3)3}&=&-2k\, \left\{{\frac { -{m}^{2}(k'+\sigma)+2\,k'\left[
 \left(\dot{\alpha}(t) \right) ^{2}+\ddot{\alpha}\left( t \right) \alpha \left( t \right)\right] }{\sqrt {k'+\sigma}}}\right\}{{\rm e}^{mx}}\,.
\end{eqnarray}

If we use the differential area element as $dS_j=(dydz,0,0)$, then the momentum flux in the (x)-direction will be

\begin{equation}
\frac{dP^{(1)}}{dt}=\left(\frac{L^2}{4\pi}\right)\,\left[\frac {  \left( \dot{\alpha}(t) \right) ^{2}+\alpha \left( t \right) \ddot{\alpha}\left( t \right)}{\sqrt {k'+\sigma}}\right]\, k'{\rm e}^{mx}\,,
\end{equation}
therefore the pressure in such a direction is

\begin{equation}
p_1(t)=\left(\frac{1}{4\pi}\right)\,\left[\frac {  \left( \dot{\alpha}(t) \right) ^{2}+\alpha \left( t \right) \ddot{\alpha}\left( t \right)}{\sqrt {k'+\sigma}\,\left( \alpha(t) \right) ^{2}}\right]\, k'{\rm e}^{mL}\,.
\end{equation}
We have considered the limit $x\rightarrow L$ in addition the the above integration. We also have used an unitary area $A=\alpha(t)L^{2}$ to calculate the pressure.

Similarly we can use $dS_j=(0,dxdz,0)$ to calculate the momentum flux in the (y)-direction, which reads

\begin{equation}
\frac{dP^{(2)}}{dt}=\left(\frac{L^2}{4\pi}\right)\,\left[\left( \dot{\alpha}(t)
 \right) ^{2}+\alpha \left( t \right) \ddot{\alpha}\left( t \right)\right]\,,
\end{equation}
thus the pressure in the (y)-direction is given by

\begin{equation}
p_2(t)=\left(\frac{1}{4\pi}\right)\,\left[\frac{\left( \dot{\alpha}(t)
 \right) ^{2}+\alpha \left( t \right) \ddot{\alpha}\left( t \right)}{\left( \alpha(t) \right) ^{2}}\right]\,.
\end{equation}
For the (z)-direction we use $dS_j=(0,0,dxdy)$ which yields

\begin{small}
\begin{equation}
\frac{dP^{(3)}}{dt}=\left[\frac{L\left({{\rm e}^{mL}}-1\right)}{8\pi m}\right]\left\{{\frac { -{m}^{2}(k'+\sigma)+2\,k'\left[
 \left(\dot{\alpha}(t) \right) ^{2}+\ddot{\alpha}\left( t \right) \alpha \left( t \right)\right] }{\sqrt {k'+\sigma}}}\right\}\,,
\end{equation}
\end{small}
as the momentum flux in this direction. Then using our unitary area we get

\begin{small}
\begin{equation}
p_3(x,t)=\left(\frac{{{\rm e}^{mL}}-1}{8\pi mL}\right)\left\{{\frac { -{m}^{2}(k'+\sigma)+2\,k'\left[
 \left(\dot{\alpha}(t) \right) ^{2}+\ddot{\alpha}\left( t \right) \alpha \left( t \right)\right] }{\sqrt {k'+\sigma}\,\left( \alpha(t) \right) ^{2}}}\right\}\,,
\end{equation}
\end{small}
for the pressure in the (z)-direction.

From field equations we have
\begin{equation}
p_2(t)=\frac{1}{8\pi}\left[\frac{m^2+\left( \dot{\alpha}(t)
 \right) ^{2}}{\left( \alpha(t) \right) ^{2}}\right]+\frac{3\sigma m^2}{32\pi k^{\prime}\,\left( \alpha(t) \right) ^{2}}-\left(1+\frac{\sigma}{k^{\prime}}\right)\,p\,,
\end{equation}
and
\begin{small}
\begin{eqnarray}
p_1(t)&=&\left\{\left[\frac{\frac{k^{\prime}}{8\pi}\left(m^2+\left( \dot{\alpha}(t)
 \right) ^{2}\right)+\frac{3\sigma m^2}{32\pi}}{\sqrt{k^{\prime}+\sigma}\, \left(\alpha(t) \right) ^{2}}\right]-\left(k^{\prime}+\sigma\right)^{1/2}\,p\right\}e^{mL}\,,\nonumber\\
p_3(t)&=&\left\{\frac{1}{32\pi}\left[\frac{4k^{\prime}\left( \dot{\alpha}(t)
 \right) ^{2}-m^2\sigma}{\sqrt{k^{\prime}+\sigma}\, \left(\alpha(t) \right) ^{2}}\right]-\left(\sqrt{k^{\prime}+\sigma}\right)\,p\right\}\left(\frac{{{\rm e}^{mL}}-1}{mL}\right)\,.\nonumber\\
\end{eqnarray}
\end{small}
Therefore we can see that the gravitational pressure is opposite to the fluid pressure and that the total pressure in such a system is not only due to the fluid pressure, i. e. the gravitational pressure plays an important role in the stability of a rotating Universe. This also means that in a scenario with expansion in the presence of the cosmological constant the gravitational pressure should be taken into account in order to understand cosmic expansion. We also conclude that G\"odel-type Universe is not isotropic since the total pressure is different in each direction.

\section{Conclusion}

In this paper we have worked with the energy-momentum flux of G\"odel-type space-time. We have derived an expression for the total pressure for each spatial direction in terms of pressure of the fluid, thus we showed the role of the gravitational pressure in such a space-time. We have supposed an open surface of integration in order to calculate the energy-momentum flux, thus the total pressure in the (x) and (y) directions is independent of the parameter of such a surface. If the parameter m is negative and if we consider $L\rightarrow\infty$, then the total pressure in the (y)-direction would be the only one which remains non-vanishing in such a context. This means that $p_2(t)$ would be the only pressure observable. We point out that the total pressure is not just the pressure of the fluid thus the gravitational pressure plays an important role in the understanding of G\"odel-type solution of Einstein equations.

\end{document}